\documentclass{article}
\usepackage[utf8]{inputenc}
\usepackage{array, makecell}
\usepackage{amsthm}
\usepackage{amssymb}
\usepackage{xspace}

\newtheorem{theorem}{Theorem}[section]
\newtheorem{definition}{Definition}[section]
\newtheorem{exdef}{Definition}
\newtheorem{ex}{Example}[section]

\newcommand{\natnrs}{\mathbb{N}}
\newcommand{\ra}{\rightarrow}

\newcommand{\voc}{\Sigma}
\newcommand{\cf}{{\mathcal{C\!F}}}

\newcommand{\Herb}[1]{{{\mathcal H}(#1)}}
\newcommand{\HerbAx}{Herbrand Axiom\xspace}

\newcommand{\Log}{{\mathcal{L}}}
\newcommand{\LogD}{{\mathcal{L}_{\mathcal D}}}
\newcommand{\modelsFO}{\models_{FO}}
\newcommand{\modelsD}{\models_{\mathcal D}}
\newcommand{\modelsL}{\models_{\Log}}

\newcommand{\sym}{\sigma}
\newcommand{\rul}{\leftarrow}
\newcommand{\defp}[1]{Def(#1)}
\newcommand{\param}[1]{Param(#1)}

\newcommand{\Tr}{{\mathbf t}}
\newcommand{\Fa}{{\mathbf f}}

\newcommand{\I}{{M}}
\newcommand{\J}{{N}}
\newcommand{\dom}[1]{{\mathcal{U}^{#1}}}

\newcommand{\lp}{\Pi} 

\newcommand{\LHM}[2]{LHM^{#1}({#2})}

\newcommand{\ignore}[1]{}

\title{ The Logic of Logic Programming\\
}
\author{Marc Denecker and David S.~Warren}
\date{October 2022}

\begin{document}

\maketitle

\begin{abstract}
Our position is that logic programming is not programming in the Horn clause sublogic of classical logic, but programming in a {\em logic of (inductive) definitions}. Thus, the similarity between prototypical Prolog programs (e.g.,  {\tt member,  append, }\dots)  and how inductive definitions are expressed in mathematical text, is not coincidental but essential. We argue here that this provides a natural solution to the main lingering semantic questions of  Logic Programming and its extensions.

\end{abstract}

\section{Introduction}

There is much ado about the ``declarative semantics''  of logic programs. But for a pure positive logic program, virtually every logic programmer around the world will accept  that the Least Herbrand Model (LHM)  \cite{jacm/EmdenK76}  represents the state of affairs of the universe and the predicates as determined by the program\footnote{Negation in programs is discussed in Section~\ref{sec-negation}.}. 
E.g., take  the member program:
\begin{verbatim}
    member(X,[X|T]).
    member(X,[H|T]) :- member(X,T). 
\end{verbatim}
Assuming the vocabulary consists of predicate member/2, list functor $|/2$ and the constant symbols [], 0, 1, 2, \dots, then the domain of the LHM is the Herbrand universe, the set of  terms built from these symbols, and the LHM contains  $member(t,l)$ iff $t$ is a member of list $l$ 
($l$ not necessarily ending with []).

Historically, e.g., in \cite{jacm/EmdenK76},  logic programs are explained as Horn theories, i.e., sets of material implications. But,  Horn theories are satisfied in an extremely  broad class of (Herbrand and first-order) structures, most of which do not match at all  with the LHM. For instance, the Herbrand structure in which all {\tt member} atoms are true, satisfies the Horn theory while it is {\em full of errors}, e.g., {\tt member(0,[1,2,3])}.  It is proven in \cite{jacm/EmdenK76} that the Horn theory logically entails all atomic facts true in the LHM. However, the Horn theory entails {\em none} of the intended negative literals (e.g., $\neg member(0,[1,2,3])$) and only few of the composite formulas true in the LHM and expected to hold in a theory of list-membership in the given universe. Arguably, the Horn theory is inadequate as an  ``axiomatisation'' of the LHM and the way programmers interpret their program.  This inadequacy emerges not only for the {\tt member} program, but for virtually all logic programs that logic programmers write.  

Logic programming is two languages in one: a declarative and a procedural one. Our goal is to investigate its declarative logic. We argue that such a logic should at least play the following roles.  (1) The logic should explain what the states of affairs are that programmers associate with their program. Here in this case, it is the LHM and only the LHM.  In this logic, a logic program should be an ``axiomatisation'' of the LHM, in the sense that mathematical logicians see this. (2) The declarative logic should be related to a range of natural language expressions that closely  correspond to the formal expressions, both in syntax and semantics.  (3) (and related) The declarative logic should give non-ambiguous insight in the meaning of the  main connectives of logic programs (``{\tt :-}'' ``{\tt ,}'' and ``{\tt not}''). (4) The declarative logic should explain not only the meaning of  finalized logic programs but also the meaning of programs  in development, components of programs. It  should explain how to interpret the program while the programmer is  writing it, and not only when it is finished. E.g., when applying logic programming to query a family  database $DB$ for siblings, the programmer may write:
\begin{verbatim}
    sibling(X,Y) :- child_of(X,P), child_of(Y,P), X \== Y.
\end{verbatim}
The programmer writes this query without knowledge of $DB$ and must have a precise understanding of this query on its own, independent of the $DB$. The declarative logic should explain this. Notice that  the LHM semantics itself,  for all its virtues, does {\em not} provide formal semantics for such stand-alone program components. 

The goal of this paper is to present such a declarative logic and argue that it satisfies the above criteria. It is based on existing ideas \cite{tocl/DeneckerBM01,tocl/DeneckerT08,KR/DeneckerV14}. A logic program is seen as a combination of definitions of all its predicates, along with an implicit ``axiom'' that constrains the set of function symbols to be the constructors of the universe.  E.g., the two rules of the {\tt member} program are read as an inductive definition of membership in mathematical text.
\begin{quote} {\bf Definition} \em We define list membership by induction:\vspace{-2mm}\begin{itemize}
    \item (base case) x is a member of list [x$|$t];
    \item  (inductive case) x is a member of  [h$|$t] if x is a member of t. 
    \end{itemize} 
    \end{quote} 
This mathematical (non-formal) definition defines  the membership  as the least relation that satisfies these two rules. Equivalently, it is the limit of the {\em induction process}: the process that starts with the empty set, and proceeds  by iterated application of these rules until a fixpoint is reached. In this view, the similarity between prototypical logic programs and (non-formal) inductive definitions in mathematical text is not coincidental but essential. As a result, the second condition of the previous paragraph will be satisfied. 

The paper is structured as follows. Section~\ref{sec-logic} introduces the logic underlying LP and discusses the meaning of the rule operator. Sections~\ref{sec-full-LP} and~\ref{sec-components-LP}  use it to analyze  full logic programs and their components. Section~\ref{sec-negation} considers negation.

\section{The declarative logic of Logic Programming}
\label{sec-logic}

As explained in the Introduction, we now design the core declarative logic $\LogD$  (syntax and formal semantics) to formalize  a logic program as a combination of an  {\em axiom of the universe}  and a {\em definition}  of all its predicates.

Most mathematical and philosophical logicians will agree that, to understand an expression, one needs to know its {\em truth conditions}: the states of affairs in which it is {\em true}, and the states of affairs in which it is {\em false}. A logic semantics that specifies this is a {\em truth conditional  semantics}. The common way to formalize  this is through a satisfaction relation $\I\models \psi$ (or a truth function $\psi^\I$). Here, $\psi$ is an expression, theory or program, and $\I$ a structure that is an abstraction of a state of affairs. First-order logic's (FO) satisfaction relation $\models_{FO}$ specifies  a truth conditional semantics.  A semantics of this type abstracts all computational and operational aspects and formalizes the answer to the essential question of declarative meaning: {\em when is an expression true, when is it false?} In our case, the expressions $\psi$ will be logic programs $\lp$ and their components: an {\em axiom of the universe} and component definitions. 
E.g.,  a structure $\I$  interpreting the relation {\tt member} by a relation containing $(0,[1,2,3])$ {\em will not satisfy} the {\tt member} definition and is not a {\em model} of the definition. Hence, this definition is {\em false} in this structure. 

The semantics of FO is based on {\em first-order structures}, while  in Logic Programming, often only {\em Herbrand structures} are used. In this paper we will use first-order structures. In particular, we cannot formalize the meaning of an axiom of the universe without considering also structures (abstractions of states of affairs) that {\em do not } satisfy it. 

\begin{definition} A (first-order) structure $\I$ for a (first-order) vocabulary $\voc$ consists of a non-empty universe $\dom{\I}$, and appropriate values $\sym^\I$ in $\dom{\I}$ for all symbols $\sym\in\voc$ (an element of, or a relation or function of appropriate arity on $\dom{\I}$). 
\end{definition}
\begin{definition} The Herbrand universe $HU(\voc)$ for  $\voc$ is the set of terms over $\voc$.
$\I$  is a Herbrand structure of $\voc$ if  $\dom{\I}$ is $HU(\voc)$  and constants $c$ and functors $f/n$  have {\em Herbrand} values, i.e.,   $c^\I=c$ and $f^\I(t_1,\dots,t_n)=f(t_1,\dots,t_n)$.
\end{definition}
This is the standard notion of Herbrand structure except when $\voc$ contains no constant symbols. In that case,  $HU(\voc)$ is empty and no Herbrand structures exist. Instead, in logic programming semantics, one will often add an arbibrary constant to $\voc$ so that  Herbrand structures always exist.

First-order structures do have one complication compared to Herbrand structures: while two different Herbrand structures represent necessarily different states of affairs, all {\em isomorphic}  first-order structures $\I, \J$ (notation $\I\cong \J$) represent the {\em same} state of affairs. The following is a natural property for any logic $\Log$ with satisfaction relation $\modelsL$:  for any pair of isomorphic structures $\I, \J$, and for any expression (theory, program) $\psi$, $\I\models \psi$ iff $\J\models \psi$. Stated differently, $\I$ is a model of $\psi$ iff $\J$  is a model of $\psi$.  We will show that our semantics  satisfies this constraint (Definition~\ref{def-H} and Theorem~\ref{theo-iso}).

\begin{definition}    For a given vocabulary $\voc$, and two (first-order) structures $\I, \J$ interpreting at least $\voc$, we say that $\I$ and $\J$ are isomorphic relative to $\voc$ (notation $\I \cong_\voc \J$) if there exists a 1-1 mapping  $b:\dom{\I}\ra\dom{\J}$ such that $b(\sym^\I)=\sym^\J$ for each  $\sym\in\voc$. \footnote{Here  $b$ is  extended to functions and relations on $\dom{\I}$ in the standard way. } Two structures $\I, \J$ are isomorphic (notation $\I\cong\J$) if they interpret the same set of symbols $\voc$ and $\I\cong_\voc\J$.
\end{definition}\vspace{-1mm}

Below, we define the syntax and the satisfaction relation $\modelsD$ of the logic $\LogD$ with a truth conditional semantics for logic programs $\lp$ and their components. The outcome should be that $\I\modelsD \lp$ iff $\I$ is (isomorphic to) the LHM of $\lp$. The logic $\LogD$ will turn out to be a sublogic of the logic FO(ID) \cite{tocl/DeneckerT08} containing definitions with a.o., negation in the body and integrating it with FO.

\vspace{-3mm}

\paragraph{The \HerbAx}

Logic Programming uses constant and function symbols in a radically more restrictive  way than in FO. In LP,   they are treated as constructors of the universe of the program; in FO, the universe is arbitrary, and constants and functors may denote arbitrary objects and functions in it. Semantically, this leads LP  to use  Herbrand structures, while FO admits the much broader class of first order structures. 

Why does Prolog consider only Herbrand structures?
To the programmer, the universe of the program consists of all data structures. Compound terms (e.g.,  the  term [1,2,3]) are used as data structures, i.e., containers of data, from which later  data can be retrieved through unification. 
But this works only when functors  are interpreted as {\em constructors}. 
For example, consider the  first-order structure $\I'$ for the {\tt member} Horn theory with universe $\{a\}$. All constants are interpreted by $a$, every n-ary function symbol is interpreted as the map of n-tuple $(a,\dots,a)$ to $a$, every n-ary predicate is interpreted as $\{(a,\dots,a)\}$. Structure $\I'$  provably  satisfies the Horn theory. 
The data structures represented by terms {\em all} collapse into the same object $a$; all information stored in them has vanished and is completely lost. This is why constructors are  important in logic programming. It is the same in functional programming. Also there,  the universe of a program is built as the collection of terms formed by a set of constructors. Of course, functional programs have many defined non-constructor functions as well. In LP, the only such functions are the interpreted ones, e.g., $+, \times$.

The  {\em \HerbAx for } a set $\cf$ of  constant and function symbols is syntactically denoted as  $\Herb{\cf}$. It expresses that the universe is $HU(\cf)$ and that symbols in $\cf$ are its {\em constructors}. 
\begin{definition}\label{def-H}
A structure $\I$ satisfies $\Herb{\cf}$, notation $\I\modelsD\Herb{\cf}$, if  $\dom{\I}$ is $HU(\cf)$ and all symbols in $\cf$ have Herbrand values. Or,  if $\I$ is isomorphic to such a  structure.
\end{definition}
Recall that if $\cf$ contains no constants, then $HU(\cf)$ is empty and  $\Herb{\cf}$ is inconsistent.

The \HerbAx can be expressed as a combination of the unique name axiom and the domain closure axiom for $\cf$.   The domain closure axiom cannot be expressed in FO but requires second-order logic or inductive definitions. More  discussion is  beyond the scope of this article.

Not every structure $\I$  satisfying  $\Herb{\cf}$ is a Herbrand structure. 
If $\I$ interprets a function symbol $f/n$ not in $\cf$, $\I$ is not a Herbrand structure and  $f/n$ is  not interpreted as a constructor.  E.g., take again $\cf= \{ [], |/2, 0, 1, 2, \dots \}$. Take its  unique Herbrand structure  and expand it for the numerical product functor $\times/2$ to the structure $\I_\times$ by interpreting  $\times$ by the function $\times^{\I_\times}: HU(\cf)^2\mapsto HU(\cf) $ that maps pairs of numbers $n, m$ to $n\times m$, the product of $n$ and $m$, and maps all other pairs to $[]$. Although $\I_\times$ is not a Herbrand structure, it satisfies $\Herb{\cf}$. Such structures are needed for the semantics of Constraint LP, e.g.,  CLP(R); they are also used in examples below.

\paragraph{The simplest definition logic}

Definitions, non-formally, are a common and precise form of human knowledge. E.g., a non-inductive  definition of sibling: 

\begin{exdef}
A person $x$ is a sibling of $y$ if \footnote{In non-formal definitions, ``if'' is often used where, logically speaking, ``iff'' is intended.} $x$ and $y$ are different and they share a parent.  
\end{exdef}

An inductive definition is that of the reachablity relation of a graph.

\begin{exdef}
    We define the reachability relation $R$ of graph $G$ by induction:
    \begin{itemize}
        \item if $(x,y)\in G$ then $(x,y)\in R$;
        \item if $(x,y)\in R$ and $ (y,z)\in G$ then $(x,z)\in R$.
    \end{itemize}   
\end{exdef}

(Non-formal) definitions {\em define}  concepts {\em in terms  of} other concepts. We call the latter the {\em parameters} of the definition. E.g., {\em sibling} is defined in terms of {\em  parent}, and the reachability relation $R$ in terms of $G$. A definition does not constrain the parameter concepts but derives, for each possible assignment of values for the parameters, a unique value for the defined concept.

Basically a (non-formal, inductive)  definition specifies how the value (or extension) of the defined set  is obtained from the value of the parameters. It can be explained in two equivalent ways: non-constructively, the defined set is the least set that satisfies the rules {\em interpreted as material implications}; constructively, the defined set is the result of the {\em induction process}: starting from the empty set, the rules are iteratively applied in a bottom up fashion until a fixpoint is reached.  The constructive and non-constructive  methods are well-known to be equivalent.

Importantly, while the above explanations are typically used  for inductive definitions, both work for non-inductive definitions as well. E.g., it may be overkill to use an induction process to construct the value of the sibling relation from a given parent relation, but it does construct the correct relation.

We now introduce the formal syntax for expressing definitions.  
\begin{definition}
A formal definition $D$ is a non-empty set of definitional rules of the form:
\[ A \rul B_1,\cdots, B_n.\]
where $A, B_1, \dots, B_n$  are standard atomic formulas.   We allow $B_i$ also to be $\Tr, \Fa$ and  equality and disequality atoms $s=t, s\neq t$. All variables are implicitly quantified universally in front of the rule. A predicate symbol is defined by $D$ if it occurs in the head of a rule; a parameter symbol of $D$ is any other non-variable symbol that occurs in $D$. The set of defined predicates is denoted  $\defp{D}$, the set of parameters as $\param{D}$.
\end{definition}
This rule-based definition construct is syntactically similar to Aczel's (propositional)  definition logic in \cite{Aczel77}  and  Martin-L\"of (predicate) definition logic in \cite{MartinLoef71}. However, following intuitionistic tradition, Martin-L\"of proposed only proof theory, not formal model semantics. 

\begin{definition} A formal definition $D$ is {\em inductive} if its dependency graph\footnote{The graph consisting of pairs $(P,Q)$ of predicate symbols such that $Q$ appears in the head and $P$ in the body of a rule in $D$.} contains a cycle. 
\end{definition}

E.g., the formal representation of the sibling definition: 
\[D_s = \left \{\begin{array}{l}
sibling(x,y) \rul  child\_of(x,z), child\_of(y,z), x\neq y
\end{array}\right\}\]

E.g., the formal representation of the non-formal definition of 
reachability: 
 
\[ D_R = \left \{ \begin{array}{l}
R(x,y) \rul G(x,y)\\
R(x,z) \rul R(x,y), G(y,z)
\end{array}\right\} \]

E.g., an inductive definition of the product of a list of numbers:
\[D_L = \left \{  \begin{array}{l}
Listproduct([],1) \rul \\
Listproduct([h|t], h\times p)\rul Listproduct(t,p)
\end{array}\right\}\]

A formal definition can define multiple predicates by simultaneous induction. For a non-formal example, take the 
inductive definition of even and odd numbers:  0 is even; if n is even then n+1 is odd; if n is odd, n+1 is even. 

\ignore{
We distinguish between the definitional implication $\rul$ and the standard material implication $\Rightarrow$. We  define:
\[ FO(A \rul B_1,\dots,B_n) =  \forall x_1, \dots ,x_n: B_1\land \dots\land B_n \Rightarrow A\]
where $x_1,\dots, x_n$ is the set of variables in the rule.  Define $FO(D) $ as the FO theory $\{FO(r) | r\in D\}$.  
}

 From now on, sets $D$ of rules can be seen as Horn theories and as definitions, and they can be evaluated in two satisfaction relations $\modelsFO$ and $\modelsD$.  

Below, we define the satisfaction relation $\modelsD$ for definitions $D$, basically by copy paste of the above non-constructive explanation of non-formal definitions. 
\begin{definition}\label{def-sat}
    Let $D$ be a definition and $\I$ a first-order structure  interpreting all symbols in $D$. We define that $\I$ satisfies $D$  (notation $\I\modelsD D$) if $\I\modelsFO D$ and for all structures $\J$  with the same universe  and values  of  parameters  in $\param{D}$  as  $\I$,  if $\J\modelsFO D$ then it holds that  $p^\I\subseteq p^\J$ for all $p\in \defp{D}$. \
\end{definition}

A definition does not in itself incorporate an assumption about the nature of constants and functors as constructors. This is why $\modelsD$ needs to be defined in terms of first order structures. Thanks to this,  it specifies the semantics of  the definition $D_l$ of $Listproduct$ which uses the non-constructor function $\times$. 

Definition~\ref{def-sat} makes use of the non-constructive characterisation of definitions. Alternatively, the value $p^\I$ of defined predicates can be characterized constructively, stating that $\I$ is to be the limit of a (potentially transfinite) induction process $\I_0, \I_1, \dots, \I_n, \I_{n+1},\dots$ that starts at the structure $\I_0$ identical to  $\I$ except that  defined predicates are interpreted by the empty set. From then on, $\I_{n+1}$ is obtained from $\I_n$ by applying one or more or all applicable rules in a bottom up way (and taking the union values for limit ordinals). In the constructive interpretation, rules specify the atomic operations of the induction process. Martin-L\"of called them {\em productions}. 
The induction process in case of programs with and without negation was formally described in \cite{KR/DeneckerV14}.

\begin{ex} The definition $D_R$ of reachability (given above) contains only predicate symbols and has no Herbrand structures but $D_R$ has infinitely many first-order models. In each model $\I$, the value $R^\I$  is the reachability relation (a.k.a. the transitive closure) of  $G^\I$.  Consider $\I$:
\[ \dom{\I} = \{ a,b,c\}, G^\I = \{(a,b),(b,a), (c,c)\}, R^\I=\{(a,a),(b,b),(c,c),(a,b),(b,a)\}\]
To verify that $\I\modelsD D_{R}$, we can verify that $R^\I$  minimally satisfies the Horn theory $D_R$ in $\I$. Alternatively, we can build  the induction process of $D_R$ in the context of $\I$ and verify that it constructs $R^\I$. The following sequence of elements of $R$ can be derived by iterated rule application:
\[ \langle (a,b), (b,a), (a,a), (b,b), (c,c) \rangle  \] 
It is well-known that  reachability cannot be expressed in FO. \footnote{\label{foot-reach}Here is a folk proof that ``$R$ is the reachability relation of $G$'' is not expressible in FO. Assume it was expressible in FO, by the FO theory $\Psi$. Choose  new constants $A, B$, consider the FO theory $\Psi\cup\{R(A,B)\}\cup\{\psi_n|n\in\natnrs_0\}$ where $\psi_n=\neg(\exists x_1,\dots, x_{n-1}:G(A,x_1)\land\dots\land G(x_i,x_{i+1})\land\dots\land G(x_{n-1},B))$ expresses that there is no path of length $n$ from $A$ to $B$. This theory is unsatisfiable since it states that $B$ is reachable from $A$ but there are no finite paths from $A$ to $B$. Therefore by the compactness theorem it has a finite subset $\Omega$ that is unsatisfiable, which is impossible since clearly, its superset $\Psi\cup\{R(A,B)\}\cup\Omega$ is satisfiable. Indeed, $\Omega$ ``forbids'' only a finite number of lengths of paths from $A$ to $B$. QED } \footnote{\label{foot-completion}Clark completion of a rule set sometimes agrees with its semantics in $\LogD$ but not  for many inductive rule sets, e.g., $D_R$. For example, adding  $(a,c),(b,c)$ to $R^\I$ yields a model of  the Clark completion in which $R^\I$ is not the reachability relation. }

\end{ex}

\begin{ex}
A small part of the (infinite) induction process of the definition  $D_L$ of {\tt Listproduct} in the context of the structure $\I_\times$ introduced before is:
\[ \langle ([],1), ([2],2), ([3,2],6), ([5,3,2],30) ,\dots  \rangle\]
\end{ex}

Two important theorems  follow (no proof provided).  Let $D$ be a definition over vocabulary $\voc$. 
\begin{theorem} \label{theo-iso} If $\I \cong_{\voc}\J$ then  $\I\modelsD D$ iff $\J\modelsD D$.
\end{theorem}
The following theorem specifies what should be a property of every logic of definitions:  given the universe and values of the parameter undefined symbols, $D$ uniquely determines the values of its defined symbols.
\begin{theorem} \label{theo-unicity} Every structure interpreting $\param{D}$  has a unique expansion for the defined symbols that satisfies $D$.  
\end{theorem}


\section{Explaining  full logic programs}\label{sec-full-LP}

The position of this paper on the declarative reading of logic programs 
boils down to the following.
\begin{quote} A logic program $\lp$ is a theory $\{\Herb{\cf}, D\}$ of the logic $\LogD$,  consisting of a definition $D$ that defines every predicate in $\lp$ and the \HerbAx $\Herb{\cf}$ for a set $\cf$ of constants and functors. 
\end{quote}

\noindent
It is well-known that each  program $\lp$  has a unique least Herbrand interpretation  $\LHM{\cf}{\lp}$, where $\cf$ contains (at least) all constant and function symbols  in $\lp$  \cite{jacm/EmdenK76}.  
Let $\voc$ be the set of all symbols in $\lp$ and $\cf$. 
\begin{theorem} 
A structure $\I$ satisfies $D$ and $\Herb{\cf}$ iff $\I\cong_{\voc} \LHM{\cf}{\lp}$. 
\footnote{This theorem reassures us that the LHM semantics is correct as a model semantics for $\LogD$. Importantly, the original paper \cite{jacm/EmdenK76} that introduced LHM did not mention definitions  and presented the LHM as the {\em denotation} of the Horn theory,  the set of entailed atomic formulas.
}
\end{theorem}
Thus, any $\voc$-structure $\I$ isomorphic with the LHM is also a model of $\lp$, moreover any extension of such $\I$  with {\em arbitrary values} for any set of additional symbols, is also a model of $\lp$. These are the only models of $\lp$.  The satisfaction relation $\modelsD$ implements the principle of isomorphism. Importantly, it also implements the natural principle  that  $\lp$ contains no information about symbols not occurring in $\lp$.   

\begin{ex}\label{ex-member} Taking $\cf=\{|/2\}$ for the {\tt member} program results in an inconsistent $\Herb{\cf}$. But, for any extended $\cf$ with at least one constant, this program determines  the correct membership relation within the intended universe. All predicates other than {\tt member} are unconstrained by it.
\end{ex}

\begin{ex}\label{ex-family}
A homework problem commonly given early in introductory Prolog courses is to define family relationships, such as sibling, grandparent, or ancestor, using just a binary child\_of relation, and to test it using the student's own family.  To define sibling, a student may  submit:
\begin{verbatim}
    sibling(X,Y) :- child_of(X,P), child_of(Y,P), X \== Y.
    child_of(tessa,david).
    child_of(jonah,david).
\end{verbatim}
The LHM is the unique state of affairs that the student had in mind.  The universe is as described by $\Herb{\{tessa,jonah,david\}}$. The definition can be interpreted non-constructively through  minimal satisfaction, or constructively through the induction process which constructs the intended relations in at most 4 steps. The rules can be interpreted as {\em productions} in the induction process. 
The facts of {\tt child\_of} behave, not as  a conjunction of true facts, but as an exhaustive enumeration involving that same set of facts. 
The  rule for {\tt sibling} behaves, not as a weak material implication, but as a necessary and sufficient condition to be a sibling.  
\end{ex}

\ignore{
\begin{theorem}
	For every structure $\I$ interpreting at least all symbols of $\lp$, the following statements are equivalent:
	\begin{enumerate}
		\item $\I$ is isomorphic modulo $\voc_D$  $\LHM{\cf}{\lp}$ potentially extended with arbitrary values for  other symbols.  
		\item $\I$ satisfies $\Herb{\cf})$ and $D$;
	\end{enumerate}
\end{theorem}
}

\section{Explaining components of logic programs} 
\label{sec-components-LP}

The previous section defines a program $\lp$ as consisting of two modules only: $D$ and $\Herb{\cf}$. A program can be large and complex, in which case it may really only be understood by its programmer(s) in a piecemeal way.  Therefore, a large program must be able to be split into a collection of natural, meaningful components which the programmer understands, and develops more or less independently of each other. The definitional view shines bright light on this.

Taking the constructive view of definitions, rules are to be viewed as productions in the induction process. We define predicates by describing how their values must be constructed from the values of parameters. The basic operations are bottom-up execution of the rules. As such, single rules  are not truth functional expressions. Satisfaction of a rule in a structure is not defined.
Of course a rule entails a material implication, but this captures only a small bit of how to understand it. 

This leads us to the following question:   what are the least components of logic programs for which it makes sense to ask the question: {\em  when is it true, when is it false}? It is not the rule, that much is clear.

The ``formulas'' of the logic, the basic truth functional components of a program are its  (sub)definitions,  i.e., rule subsets of  $D$ that define one or more predicates.  Take the family program of Example~\ref{ex-family}. To the programmer, the program clearly consists of three components: the (implicit) \HerbAx, and the definitions of {\tt child\_of} and of {\tt sibling}.  The first definition  defines {\tt child\_of} by exhaustive enumeration and contains no knowledge of {\tt sibling}. The second defines {\tt sibling} in terms of the parameter {\tt child\_of} and  contains no knowledge of the value of this parameter. 

Not every partition $D_1,\dots, D_n$ of $D$ yields a sensible modularization  in subdefinitions of it. A subdefinition $D_i$ that defines one or more predicates, should contain all rules of $D$ involved in the induction process of these predicates. Thus, formally, a component definition $D_i$ of $D$ should contain all rules of $D$ with the same predicates in the head. For the same reason, if two or more predicates are defined by simultaneous induction, their rules should not be spread out over multiple subdefinitions but be concentrated in one definition; otherwise, the induction process cannot be computed. Thus, there should not be cycles in the dependency relation over multiple subdefinitions. 
In summary,  the definitional view on logic programming suggests that  a program $\lp$ can be naturally split in modules $D_1,\dots, D_n,\Herb{\cf}$, such that  each predicate is defined in exactly one module $D_i$ and there are no cycles in the dependency relation involving predicates defined in different modules. The following theorem proves the correctness of this hypothesis. 

\begin{theorem} 
For every structure $\I$ interpreting all symbols of $\lp$ and $\cf$, the following statements are equivalent:
\begin{enumerate}
	\item $\I \cong_{\voc_D} \LHM{\cf}{\lp}$ 
    \item $\I$ satisfies  $D$ and $\Herb{\cf})$;
    \item $\I$ satisfies  $D_1, \dots, D_n$ and  $\Herb{\cf})$. 
\end{enumerate}
\end{theorem}

\ignore{
\begin{proof} [Sketch] Take any subset $S$ of $\{ D_1,\dots,D_n\}$ with the property that for every element $D_j\in S$, for every  $D_i\prec D_j$, it holds that $D_i \in S$. Then $S$  defines all predicates that appear in it, so it is a logic program. The induction hypothesis is that for every such $S$, for every first order structure $\I$ interpreting $S$ and $\cf$, the theorem holds.

Base case, let $S=\emptyset$. Then $\voc_S=\cf$ and $\I \cong_{\voc_S} \LHM{\cf}{S}$ iff $S$ satisfies $\Herb{\cf}$.

Inductive case: assume the theorem is satisfied for $S$ and let $D_i$ be lowest definition in $\prec$ not in $S$. We prove that the theorem holds for $S'=S\cup\{D_i\}$. But this is an easy consequence of Theorems~\ref{theo-iso} and \ref{theo-unicity}. {\tt Yeah yeah! CHeck }
\end{proof}
}

A proof of this theorem can be found in \cite{tocl/DeneckerT08}.
The theorem says that  $\lp$ is logically equivalent to $\Herb{\cf}$ and  the conjunction of the definitions $D_1,\dots,D_n$. This indeed shows that a programmer can develop such modules $D_i$ and reason with them independently of each other.

\section{Negation} \label{sec-negation}

The nature of negation (as failure) is probably  the most troubling question in the history of Logic Programming.\footnote{The LP community seems much less worried about the nature of the rule operator, while for more than hundred years, the logic science knows there are considerable troubles with material implication. }  For 50 years now, the general conviction is that the negation {\tt not} in Prolog cannot be classical negation $\neg$. Where does that idea come from?  It comes from the fact that Horn theories do not  entail the falsity of atoms.  E.g., consider the query
\begin{verbatim}
     ?- member(0,[1,2,3])
     no
\end{verbatim}
The answer ``{\tt no}'' expresses that the  member Horn logic theory  does not entail the truth of {\tt member(0,[1,2,3])} but neither does it entail its falsity.  As a consequence, Prolog would be unsound if in the following query, the symbol {\tt not} is interpreted as classical negation $\neg$:
\begin{verbatim}
     ?- not member(0,[1,2,3])
     yes
\end{verbatim}
This  is the first and main reason why {\tt not} is believed to be non-classical negation.

But the definitional view sheds a completely different light on the issue. Definitions augmented with the \HerbAx do indeed semantically entail the falsity of many defined atomic facts. In particular, since the LHM is the unique model (modulo isomorphism), any defined atom $A$ that is false in the LHM is false  in {\em every} model. Therefore,  this theory semantically entails the truth of the classically negated fact $\neg A$.  In particular, {\tt member(0,[1,2,3])} is false in {\em every} structure that satisfies the member program (as defined in this paper), and hence, $\neg${\tt member(0,[1,2,3])} is semantically entailed by the program.

And so, in the following procedure {\tt compress/2} which removes duplicates in a list, {\tt not} can and should be interpreted as classical negation $\neg$:
\begin{verbatim}
    compress([],[]).
    compress([X|T],[X|T1]) :- compress(T,T1), not member(X,T1). 
    compress([X|T],T1) :- compress(T,T1), member(X,T1). 
\end{verbatim}
The resulting program (defining {\tt compress} and {\tt member}) is a stratified program.  

{\em Stratification} is also  a natural principle of great importance in mathematics and science: it is the principle  that once a  concept is well-defined, it can be used to define new concepts  in terms of it. It is a key building principle of science. Mathematical logicians have studied infinite, even transfinite stratified stacks of definitions, called {\em iterated inductive definitions} \cite{Feferman70,MartinLoef71}. The well-known principle of definition by structural induction is an instance of this \cite{KR/DeneckerV14}. In \cite{tocl/DeneckerBM01,tocl/DeneckerT08,KR/DeneckerV14}, it was argued that the well-founded semantics implements a semantic stratification principle, and that logic programs under this semantics can be viewed as a finite description of this type of definition.

To conclude, the definitional view on logic programs sheds a different light on the nature of  language constructs: negation as failure is indeed classical negation! It is the rule operator that is non-classical: much stronger than a material implication, it is a production operator in the induction process. 

We conclude by noting that we have realized the four goals put forward in the Introduction.

\bibliographystyle{plain}

\begin{thebibliography}{1}

\bibitem{Aczel77}
Peter Aczel.
\newblock An introduction to inductive definitions.
\newblock In J.~Barwise, editor, {\em Handbook of Mathematical Logic}, pages
  739--782. North-Holland Publishing Company, 1977.

\bibitem{tocl/DeneckerBM01}
Marc Denecker, Maurice Bruynooghe, and Victor Marek.
\newblock Logic programming revisited: Logic programs as inductive definitions.
\newblock {\em ACM Trans. Comput. Log.}, 2(4):623--654, 2001.

\bibitem{tocl/DeneckerT08}
Marc Denecker and Eugenia Ternovska.
\newblock A logic of nonmonotone inductive definitions.
\newblock {\em ACM Trans. Comput. Log.}, 9(2):14:1--14:52, April 2008.

\bibitem{KR/DeneckerV14}
Marc Denecker and Joost Vennekens.
\newblock The well-founded semantics is the principle of inductive definition,
  revisited.
\newblock In Chitta Baral, Giuseppe {De Giacomo}, and Thomas Eiter, editors,
  {\em {KR}}, pages 1--10. AAAI Press, 2014.

\bibitem{Feferman70}
Solomon Feferman.
\newblock {Formal theories for transfinite iterations of generalised inductive
  definitions and some subsystems of analysis}.
\newblock In A.~Kino, J.~Myhill, and R.E. Vesley, editors, {\em {Intuitionism
  and Proof theory}}, pages 303--326. North Holland, 1970.

\bibitem{MartinLoef71}
Per {Martin-L\"{o}f}.
\newblock {Hauptsatz for the intuitionistic theory of iterated inductive
  definitions}.
\newblock In J.E. Fenstad, editor, {\em Second Scandinavian Logic Symposium},
  pages 179--216, 1971.

\bibitem{jacm/EmdenK76}
Maarten~H. {van Emden} and Robert~A. Kowalski.
\newblock The semantics of predicate logic as a programming language.
\newblock {\em J. ACM}, 23(4):733--742, 1976.

\end{thebibliography}

\end{document}